\documentclass[journal]{IEEEtran}
\usepackage{comment}
\usepackage{amsmath}
\usepackage{amsfonts}
\usepackage[english]{babel}
\usepackage{amsthm}
\usepackage{amssymb}
\usepackage{graphicx}
\usepackage{latexsym}
\usepackage{graphicx}
\usepackage{lipsum}
\usepackage{xcolor}
\usepackage{enumitem}
\usepackage{hyperref}
\usepackage{cleveref}
\usepackage{algorithmic,algorithm}
\usepackage{orcidlink}

\title{Elevation-Aware Supplementary Uplink for Direct Satellite-to-Device Communications \vspace{4mm}}

\author{
	Rajan Shrestha$^{_{\orcidlink{0000-0003-3146-7829}}}$,~\IEEEmembership{Graduate Student Member,~IEEE}, and 
	Hayder Al-Hraishawi$^{_{\orcidlink{0000-0002-0977-9984}}}$,~\IEEEmembership{Senior Member,~IEEE}
	\thanks{
		The authors are with the Department of Electrical Engineering, 
		University of South Florida, Tampa, FL 33620 USA (email: rajanshrestha@usf.edu; hayder@usf.edu). Corresponding author: \emph{Hayder Al-Hraishawi}.
	}
}

\begin{document}
\pagestyle{empty}
\maketitle
\bstctlcite{BSTcontrol}

\begin{abstract}
Direct satellite-to-device (DS2D) communication enables standard mobile devices to connect directly to low Earth orbit (LEO) satellites, providing global coverage without reliance on terrestrial infrastructure. However, the DS2D uplink is fundamentally constrained by long propagation distances, severe path loss, and stringent user equipment (UE) power limits, making uplink reliability particularly challenging at low elevation angles and beam edges. This paper investigates the integration of supplementary uplink (SUL) technology into DS2D systems to enhance uplink robustness while preserving UE power efficiency. Leveraging the predictable geometry of LEO satellite orbits, we develop an elevation-aware SUL framework that adapts uplink operation across frequency bands based on elevation-dependent link margin estimates. The proposed approach schedules the UE to transmit on either a primary uplink carrier or a lower-frequency SUL carrier.  An elevation-aware SUL activation algorithm with hysteresis is introduced to guide uplink carrier selection while preventing frequent switching. Simulation results demonstrate that the proposed SUL framework extends effective uplink coverage toward low-elevation and beam-edge regions, improves uplink availability over a satellite pass, and achieves stable operation with a minimal number of uplink transitions under realistic UE power constraints.
\end{abstract}

\begin{IEEEkeywords}
	Direct satellite-to-device (DS2D), supplementary uplink (SUL), low Earth orbit (LEO) satellites, non-terrestrial networks (NTNs), elevation-aware carrier selection, geometry-aware switching, link margin adaptation.
\end{IEEEkeywords} 

\section{Introduction}\label{sec:intro}

Direct satellite-to-device (DS2D) communication is emerging as a transformative technology for global connectivity, enabling standard mobile devices to communicate directly with low Earth orbit (LEO) satellites without reliance on terrestrial infrastructure. This capability enhances service continuity and resilience for handheld devices across diverse scenarios, including remote, maritime, and aeronautical environments \cite{Tuzi2023}. Recent commercial deployments by Starlink, AST SpaceMobile, and Lynk Global demonstrate the practical viability of this paradigm, with several smartphone manufacturers announcing DS2D-capable devices for emergency and extended-coverage connectivity services \cite{Garcia2025}.

However, DS2D uplink transmission faces fundamental radio-frequency (RF) challenges. The long propagation distances to LEO satellites (slant ranges of approximately 600--2000 km), combined with severe free-space path loss (on the order of 160--190 dB), atmospheric attenuation, and stringent user equipment (UE) power constraints (typically limited to 23 dBm), create a critical bottleneck \cite{Bakhsh2025}. Unlike conventional satellite terminals equipped with high-gain directional antennas and high-power amplifiers, DS2D relies on omnidirectional smartphone antennas with limited transmit power, making reliable uplink transmission particularly challenging near beam edges and at low elevation angles \cite{Xu2024}. Moreover, higher-frequency bands (e.g., Ku/Ka-band), while offering larger bandwidths, suffer from increased path loss and rain attenuation, and the high orbital velocity of LEO satellites (approximately 7.5 km/s) introduces substantial Doppler shifts that scale linearly with carrier frequency \cite{NGSO_survey}.

To address these uplink limitations, this paper investigates the integration of \emph{supplementary uplink (SUL)} technology into DS2D systems. Originally introduced in the 3rd Generation Partnership Project (3GPP) Release~15 specifications for terrestrial 5G networks \cite{Rel15SUL}, SUL allows the network to configure a UE with both a primary uplink (PUL) carrier, typically operating at a higher frequency for improved spectral efficiency, and an SUL carrier at a lower frequency to enhance coverage and reliability \cite{Rinaldi2021}. In the DS2D context, SUL exploits lower-frequency uplink bands (e.g., L- or S-band, 1–4 GHz) to benefit from more favorable propagation conditions, including reduced path loss and improved robustness to shadowing and atmospheric attenuation. As a result, SUL can sustain reliable uplink transmission at low elevation angles and near beam edges when the high-frequency PUL is unable to satisfy the target signal-to-noise ratio (SNR).

Unlike carrier aggregation or multi-connectivity techniques that rely on simultaneous multi-carrier transmission \cite{Kibria2022,MaDS2D2025}, the proposed framework schedules uplink transmission on either the PUL or the SUL carrier at any given time, thereby avoiding concurrent transmissions. This \emph{either-or} operation prevents overlapping uplink signaling, ensuring compliance with UE transmit power constraints and simplifying RF front-end design, while still enabling seamless uplink coverage extension.
This paper makes the following contributions:
\begin{itemize}   
    \item We propose an elevation-aware SUL activation algorithm that exploits the deterministic geometry of LEO orbits to switch uplink carriers based on predicted link margin, while incorporating hysteresis to prevent frequent carrier switching.

    \item Through performance evaluation under representative DS2D scenarios, we demonstrate that SUL substantially extends effective uplink coverage and improves uplink availability at low elevation angles, while preserving UE power constraints with minimal switching overhead.
\end{itemize}
The remainder of this paper is organized as follows: Section~\ref{sec:system_model} presents the system model and link budget analysis. Section~\ref{sec:sul_framework} describes the proposed framework with the SUL activation algorithm. Section~\ref{sec:performance} evaluates performance through simulation, 
and Section~\ref{sec:conclusion} concludes the paper.

\begin{figure}
    \centering
    \includegraphics[width=0.95\linewidth]{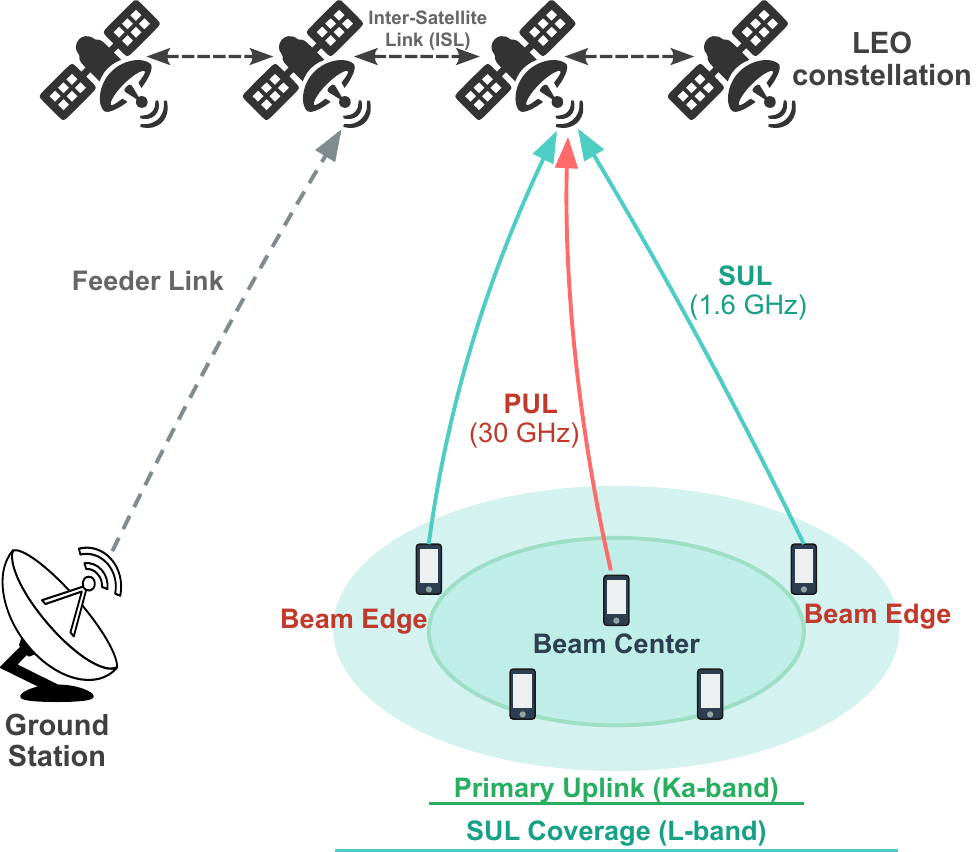}
    \caption{Schematic diagram of a LEO-based DS2D system showing PUL and SUL coverage footprints.}
    \label{fig:system_model} \vspace{-3mm}
\end{figure}
\section{System Model}\label{sec:system_model}
We consider the DS2D uplink from ground UEs to a constellation of LEO satellites equipped with regenerative payloads performing onboard baseband processing and forming multiple beams over the Earth’s surface \cite{3GPP38811v15}. Let $\mathcal{U}=\{u_1,\ldots,u_M\}$ denote the set of UEs and $\mathcal{S}=\{s_1,\ldots,s_N\}$ the set of satellites. Each UE maintains an uplink connection to a serving satellite using either a PUL carrier at frequency $f_p$ or an SUL carrier at frequency $f_s$, with $f_s < f_p$, where the lower-frequency SUL improves uplink coverage and reliability under power-limited UE constraints.

The LEO satellites operate at altitudes $h$ above the Earth’s surface. For a UE at elevation angle $\theta$, the slant range between the UE and the satellite is
\begin{equation}\label{eq:slant_range}
d(\theta)=\sqrt{(R_E+h)^2 - R_E^2\cos^2\theta}-R_E\sin\theta,
\end{equation}
where $R_E=6371$~km is the Earth’s radius. For a representative LEO altitude of $h=600$~km, the slant range varies from $600$~km at zenith to approximately $1930$km at a minimum elevation angle of $10^\circ$\cite{Wang2026}.

\subsection{Uplink Channel Model}
The DS2D uplink channel is dominated by line-of-sight (LoS) propagation, with frequency-dependent free-space path loss and atmospheric attenuation. 
The received uplink power at the satellite, expressed in dBm, is modeled as
\begin{equation}\label{eq:received_power}
\begin{aligned}
P_r(f,\theta)
&=
P_t
+ G_{\mathrm{UE}}
+ G_{\mathrm{sat}}(\theta) \\
&\quad
- PL_{\mathrm{FS}}(f,d)
- L_{\mathrm{atm}}(f,\theta)
- L_{\mathrm{impl}},
\end{aligned}
\end{equation}
where $f \in \{f_p,f_s\}$ denotes the uplink carrier frequency corresponding to the PUL and SUL, respectively. 
Here, $P_t$ is the UE transmit power (dBm), $G_{\mathrm{UE}}$ and $G_{\mathrm{sat}}(\theta)$ denote the UE and satellite antenna gains (dBi), and $L_{\mathrm{impl}}$ accounts for implementation losses (dB) such as tracking and polarization mismatch. Specifically, the free-space path loss is given by
\begin{equation}\label{eq:FSPL}
PL_{\mathrm{FS}}(f,d)=32.45 + 20\log_{10}(f_{\mathrm{MHz}}) + 20\log_{10}(d_{\mathrm{km}}),
\end{equation}
where $f_{\mathrm{MHz}}$ and $d_{\mathrm{km}}$ denote the carrier frequency and slant range in MHz and kilometers, respectively. 
The atmospheric loss term $L_{\mathrm{atm}}(f,\theta)$ accounts for gaseous absorption and rain attenuation, which increase with carrier frequency and decreasing elevation angle, following established ITU-R propagation models~\cite{ITU_P676,ITU_P618}.
The satellite antenna gain varies with off-nadir angle according to \vspace{-3mm}
\begin{equation}
G_{\mathrm{sat}}(\theta) = G_{\mathrm{sat},\max}
- 12\left(\frac{\theta_{\mathrm{off}}(\theta)}{\theta_{3\mathrm{dB}}}\right)^2, \vspace{-1mm}
\end{equation}
where $\theta_{\mathrm{off}}(\theta)$ is the off-boresight angle corresponding to elevation angle $\theta$, $\theta_{3\mathrm{dB}}$ is the 3-dB beamwidth, and $G_{\mathrm{sat},\max}$ is the peak antenna gain.

Due to the high orbital velocity of LEO satellites, the uplink signal experiences a Doppler shift given by \vspace{-1mm}
\begin{equation}\label{eq:doppler_shift}
f_D=\frac{v}{c}f\cos\phi, \vspace{-1mm}
\end{equation}
where $v\approx7.5$~km/s is the satellite velocity, $c$ is the speed of light, and $\phi$ denotes the angle between the satellite velocity vector and the LoS direction. 
The resulting Doppler shift scales linearly with carrier frequency, reaching up to tens of kHz at L/S-band and several hundred kHz at Ka-band, thereby imposing increasingly stringent frequency tracking requirements at higher frequencies.

\subsection{Uplink SNR}
The uplink SNR at the satellite receiver is given by
\begin{equation} \label{eq:SNR}
\text{SNR}(\text{dB}) = P_r(f,\theta) - \left( N_0 + 10\log_{10}(B) \right), \vspace{-2mm}
\end{equation}
where $P_r(f,\theta)$ is the received power defined in \eqref{eq:received_power}, $B$ is the channel bandwidth (Hz),
and $N_0$ is the noise power spectral density (dBm/Hz), computed as
\vspace{-2mm}
\begin{equation}
N_0 = -174 + 10\log_{10}\!\left(\frac{T_{\text{sys}}}{290}\right). \vspace{-2mm}
\end{equation}
Here, $T_{\text{sys}}$ denotes the satellite receiver system noise temperature, accounting for both antenna and receiver noise.

\section{Proposed SUL Framework}
\label{sec:sul_framework}
In LEO constellations, satellite motion and orbital geometry are highly predictable. Satellite ephemeris information enables accurate estimation of the UE elevation angle, from which key channel parameters such as slant range, free-space path loss, and Doppler shift can be inferred. Thus, uplink channel quality can be anticipated as a deterministic function of elevation angle, enabling geometry-aware uplink adaptation.
Building on this observation, this section presents an SUL framework for DS2D communications that adapts uplink operation across frequency bands based on satellite geometry to sustain reliable uplink performance for power-limited UEs. The key enabler of the proposed framework is an elevation-aware uplink carrier selection strategy that uses predicted link margin estimates to guide switching between PUL and SUL operation. At any given time, the UE is scheduled to transmit on either the PUL or the SUL, but not both simultaneously, preserving UE power efficiency and simplifying RF front-end design.

Let the predicted uplink SNR on carrier $f\in\{f_p,f_s\}$ be \vspace{-1mm}
\begin{equation}
\label{eq:predicted_snr}
\begin{aligned}
\widehat{\text{SNR}}_f(\theta)
&= \! P_t + G_{UE} + G_{sat}(\theta)
   - \! PL_{FS}\!\left(f,d(\theta)\right) \\
  &\quad - \widehat{L}_{atm}(f,\theta) 
 - N_0 - 10\log_{10}(B) - L_{\text{impl}},
\end{aligned}
\end{equation}
where $d(\theta)$ follows the slant-range model in \eqref{eq:slant_range}, 
$\widehat{L}_{atm}(f,\theta)$ denotes a nominal atmospheric and rain attenuation model, and
$G_{sat}(\theta)$ captures beam-edge gain reduction.
The corresponding predicted link margin is defined as
\begin{equation}
\label{eq:predicted_margin}
\widehat{M}_f(\theta)=\widehat{\text{SNR}}_f(\theta)-\text{SNR}_{\text{req}},
\end{equation}
where $\text{SNR}_{\text{req}}$ is the minimum SNR required for reliable uplink transmission.

Based on these margin estimates, an elevation-aware SUL activation algorithm is developed to adapt uplink operation, as summarized in Algorithm~\ref{alg:predictive_sul}. The algorithm incorporates a safety margin $\Delta_s$ to absorb modeling uncertainties (e.g., atmospheric variability and beam-edge effects) and a hysteresis margin $\Delta_h$ to prevent frequent or oscillatory carrier switching. This design enables robust DS2D uplink operation across the satellite pass while respecting UE power and hardware constraints.

\begin{algorithm}[t]
\caption{Elevation-Aware SUL Activation for DS2D}
\label{alg:predictive_sul}
\begin{algorithmic}[1]
\REQUIRE Estimated elevation angle $\theta$ (from satellite ephemeris)
\REQUIRE Current carrier $C_{active}\in\{\mathrm{PUL},\mathrm{SUL}\}$
\REQUIRE Thresholds: safety margin $\Delta_s$ (dB), hysteresis margin $\Delta_h$ (dB)
\REQUIRE Required SNR: $\text{SNR}_{\text{req}}$

\STATE Compute predicted margins $\widehat{M}_p(\theta)$ and $\widehat{M}_s(\theta)$

\IF{$C_{active}=\mathrm{PUL}$}
    \IF{$\widehat{M}_p(\theta) < \Delta_s$ \textbf{and} $\widehat{M}_s(\theta) > 0$}
        \STATE $C_{new}\gets \mathrm{SUL}$
    \ELSE
        \STATE $C_{new}\gets \mathrm{PUL}$
    \ENDIF
\ELSIF{$C_{active}=\mathrm{SUL}$}
    \IF{$\widehat{M}_p(\theta) > \Delta_s+\Delta_h$}
        \STATE $C_{new}\gets \mathrm{PUL}$
    \ELSE
        \STATE $C_{new}\gets \mathrm{SUL}$
    \ENDIF
\ENDIF
\IF{$C_{new}\neq C_{active}$}
    \STATE Execute carrier switch and update configuration
    \STATE $C_{active}\gets C_{new}$
\ENDIF
\RETURN $C_{active}$
\end{algorithmic}
\end{algorithm}

\section{Performance Evaluation}
\label{sec:performance}
\begin{table}[t]
	\centering
	\caption{Simulation Parameters for DS2D Uplink Evaluation}
	\label{tab:ds2d_params}
	\begin{tabular}{|l|c|c|}
		\hline
		\textbf{Parameter} & \textbf{SUL (L-band)} & \textbf{PUL (Ka-band)} \\
		\hline
		Carrier frequency & 1.6 GHz & 30 GHz \\
		UE transmit power & 23 dBm & 23 dBm \\
		UE antenna gain & 0 dBi & 0 dBi \\
		Satellite  antenna gain & 45 dBi & 65 dBi \\
		Minimum elevation angle & 10$^\circ$ & 10$^\circ$ \\
		System noise temperature & 290 K & 500 K \\
		Uplink bandwidth & 10 MHz & 10 MHz \\
		Atmospheric loss & 1 dB & 15 dB \\
		Implementation loss & 2 dB & 2 dB  \\
		\hline
	\end{tabular}
\end{table}
In this section, we evaluate the performance of the proposed elevation-aware SUL algorithm under DS2D operating conditions. The objective is to quantify the uplink robustness and coverage gains enabled by SUL relative to PUL-only operation. We consider a LEO satellite at an altitude of $h=600$~km serving ground UEs over a satellite pass, with a maximum UE transmit power of 23~dBm. The uplink bandwidth is fixed at 10~MHz for both the PUL and SUL carriers. The PUL operates in the Ka-band (30~GHz), while the SUL operates in the L-band (1.6~GHz). Satellite antenna gains, atmospheric losses, and system noise temperatures follow the parameters summarized in Table~\ref{tab:ds2d_params}. UE elevation angles vary from $10^\circ$ to $90^\circ$ over the satellite pass.

\begin{figure}
    \centering
    \includegraphics[width=0.95\linewidth]{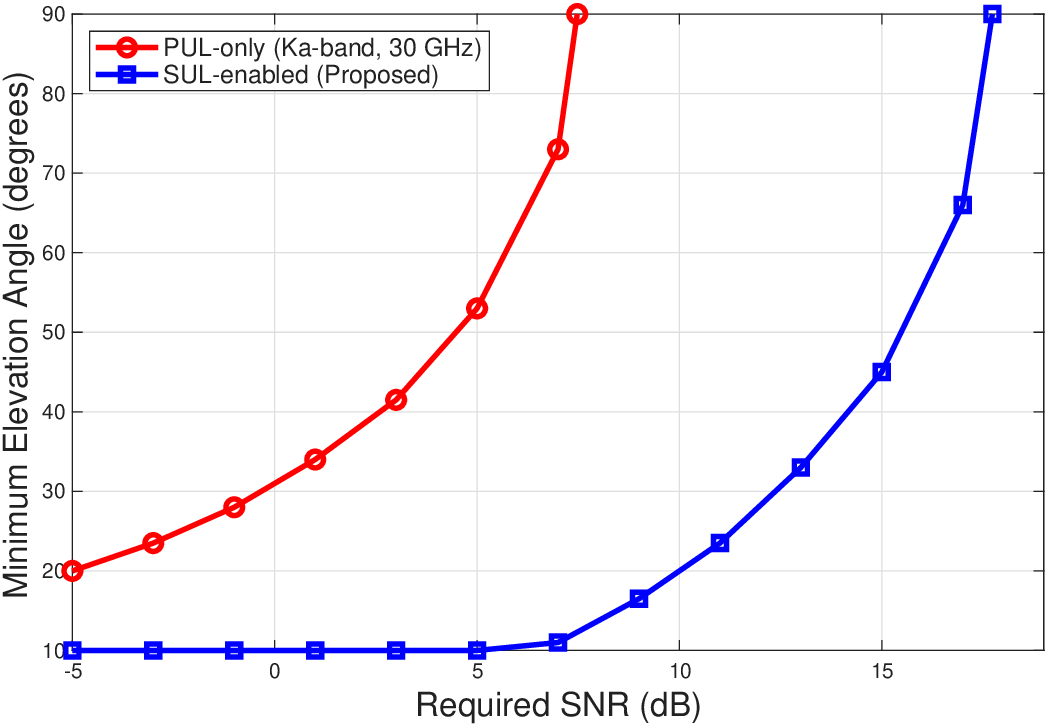}
    \caption{Minimum elevation angle required versus target SNR.}
    \label{fig1}
\end{figure}
Fig.~\ref{fig1} shows the minimum elevation angle required to sustain uplink communication as a function of the target SNR. Under PUL-only operation, the required elevation increases rapidly with the SNR target, leading to a pronounced loss of uplink coverage. In contrast, the proposed SUL-enabled framework maintains connectivity at substantially lower elevation angles across a wide range of SNR requirements. For moderate SNR targets (e.g., $0$--$5$~dB), SUL enables uplink operation close to the minimum elevation limit of the satellite footprint (around $10^\circ$), whereas PUL-only operation requires elevations above $30^\circ$--$40^\circ$. This corresponds to an elevation coverage extension exceeding $20^\circ$, demonstrating the effectiveness of SUL in supporting uplink communication near beam edges and at low elevations.

\begin{figure}
    \centering
    \includegraphics[width=0.95\linewidth]{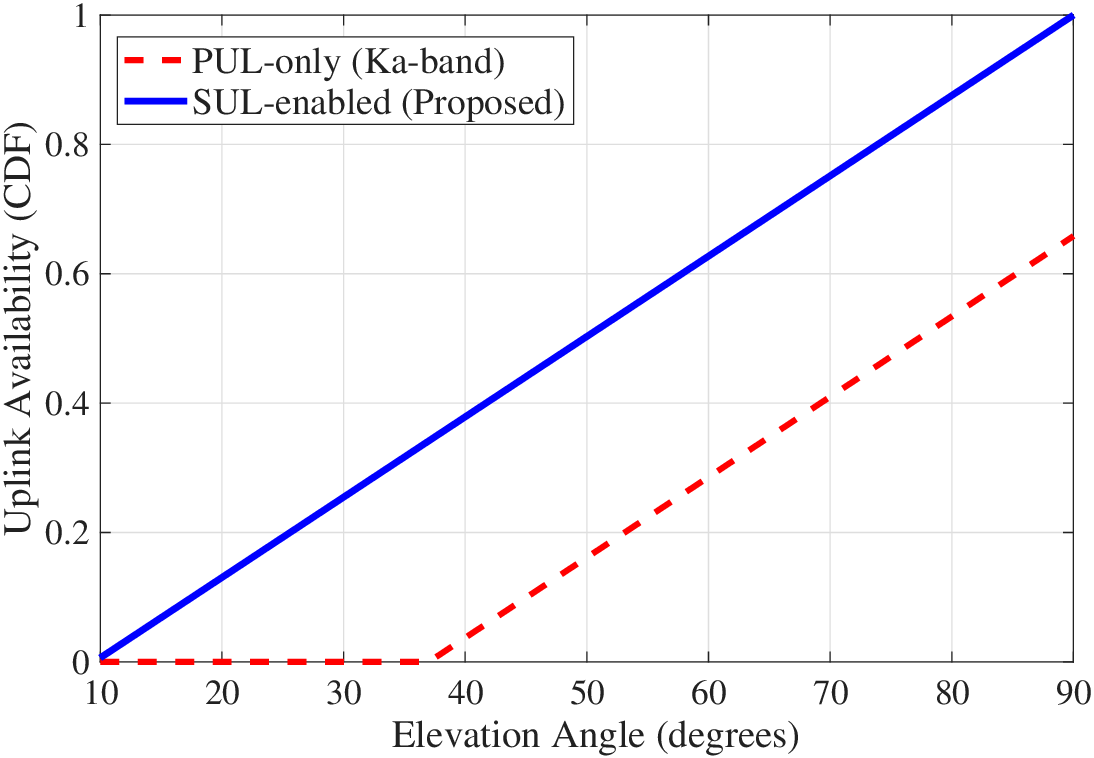}
    \caption{CDF of DS2D uplink availability versus elevation angle.}
    \label{fig2}
\end{figure}

Fig.~\ref{fig2} further quantifies this gain by showing the CDF of uplink availability across elevation angles. With SUL enabled, uplink connectivity is available from the lowest considered elevation angles, resulting in a near-uniform accumulation of availability across the satellite pass. In contrast, PUL-only operation exhibits a delayed onset of availability, becoming viable only at higher elevations where path loss and atmospheric attenuation are sufficiently reduced. Thus, the SUL-enabled framework substantially increases uplink availability across elevation angles, complementing the spatial coverage gains observed in Fig.~\ref{fig1}.

\begin{figure}
    \centering
    \includegraphics[width=0.95\linewidth]{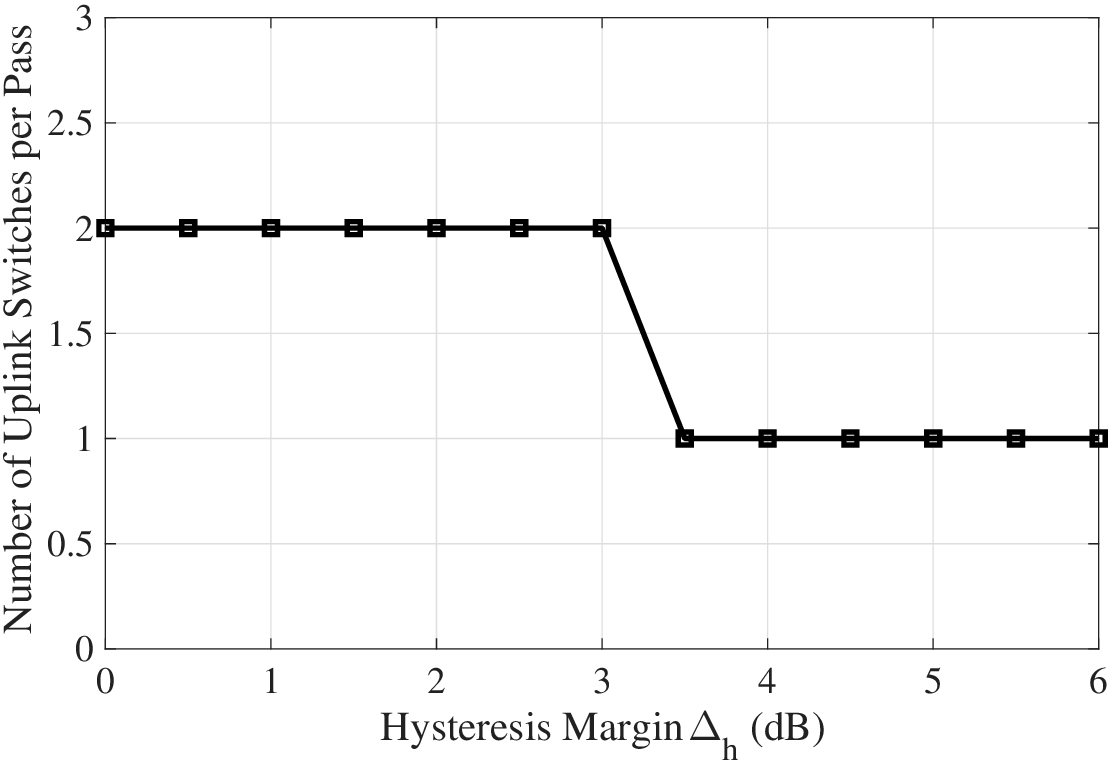}
    \caption{Number of uplink carrier switches per satellite pass versus hysteresis margin.}
        \label{fig3}
\end{figure}

Finally, we evaluate the stability of the proposed SUL framework by examining the number of uplink carrier switches during a satellite pass. Fig.~\ref{fig3} shows the number of uplink carrier switches per pass as a function of the hysteresis margin $\Delta_h$. For small hysteresis margins, the uplink may switch multiple times between PUL and SUL as the predicted link margin fluctuates around the switching threshold. However, once the hysteresis margin exceeds approximately $3.5$~dB, the proposed SUL framework consistently requires only a single uplink transition per satellite pass. This behavior demonstrates stable carrier selection and effectively eliminates ping-pong switching, confirming that the proposed geometry-aware hysteresis mechanism achieves a favorable tradeoff between uplink robustness and switching stability.

\section{Conclusion}\label{sec:conclusion}
This paper proposed an SUL framework for DS2D communications aimed at improving uplink robustness under low-elevation and beam-edge operating conditions. By leveraging a lower-frequency SUL carrier in conjunction with a geometry-aware switching mechanism, the proposed framework mitigates the severe path loss and atmospheric attenuation limitations inherent to high-frequency PUL operation. Simulation results demonstrated that the SUL-enabled scheme substantially extends the effective uplink coverage region compared to PUL-only operation, enabling reliable communication at lower elevation angles across a wide range of SNR requirements. The proposed approach also significantly increases uplink availability over a satellite pass, improving connectivity from early visibility points without violating UE power constraints. 
Overall, these results highlight SUL as a practical and low-complexity enhancement for DS2D uplink design, particularly in challenging propagation environments.



\bibliographystyle{IEEEtran}
\bibliography{IEEEabrv,References}

\end{document}